\documentclass[a4paper,11pt]{article}
\usepackage{jheppub} 
\pdfoutput=1
\usepackage{amsmath,amssymb,amsfonts}
\usepackage{graphicx}
\usepackage[normalem]{ulem}
\usepackage{slashed}
\usepackage{color}
\usepackage{braket}
\usepackage{tikz}
\usepackage{url}
\usepackage{subcaption}

\usepackage{hyperref}
\newcounter{qnumber}

\newcommand{\etr}[3]{\left\langle #1,#2,#3\right\rangle}

\newcommand{\beq}{\begin{equation}}
\newcommand{\eeq}{\end{equation}}

\usepackage{stackrel}

\makeatletter\g@addto@macro\bfseries{\boldmath}\makeatother
\DeclareMathSymbol{\shortminus}{\mathbin}{AMSa}{"39}
\def\figureautorefname~#1\null{fig.\,#1\null}
\def\equationautorefname~#1\null{eq.\,(#1)\null}

\makeatletter
\DeclareFontFamily{OMX}{MnSymbolE}{}
\DeclareSymbolFont{MnLargeSymbols}{OMX}{MnSymbolE}{m}{n}
\SetSymbolFont{MnLargeSymbols}{bold}{OMX}{MnSymbolE}{b}{n}
\DeclareFontShape{OMX}{MnSymbolE}{m}{n}{
    <-6>  MnSymbolE5
   <6-7>  MnSymbolE6
   <7-8>  MnSymbolE7
   <8-9>  MnSymbolE8
   <9-10> MnSymbolE9
  <10-12> MnSymbolE10
  <12->   MnSymbolE12
}{}
\DeclareFontShape{OMX}{MnSymbolE}{b}{n}{
    <-6>  MnSymbolE-Bold5
   <6-7>  MnSymbolE-Bold6
   <7-8>  MnSymbolE-Bold7
   <8-9>  MnSymbolE-Bold8
   <9-10> MnSymbolE-Bold9
  <10-12> MnSymbolE-Bold10
  <12->   MnSymbolE-Bold12
}{}

\let\llangle\@undefined
\let\rrangle\@undefined
\DeclareMathDelimiter{\llangle}{\mathopen}%
                     {MnLargeSymbols}{'164}{MnLargeSymbols}{'164}
\DeclareMathDelimiter{\rrangle}{\mathclose}%
                     {MnLargeSymbols}{'171}{MnLargeSymbols}{'171}
\makeatother

\begin{document}

\title{Non-Abelian and Type-A Conformal Anomalies from Euler Descent}
\author[a]{Gleb Aminov,}
\affiliation[a]{Racah Institute of Physics, Hebrew University of Jerusalem, Jerusalem 91904, Israel}
\emailAdd{gleb.aminov@mail.huji.ac.il
}

\author[b]{Csaba Cs\'aki,}
\affiliation[b]{Department of Physics, LEPP, Cornell University, Ithaca, NY 14853, USA}
\emailAdd{csaki@cornell.edu}

\author[a]{Ofri Telem,}
\emailAdd{t10ofrit@gmail.com}

\author[c]{and Shimon Yankielowicz}
\affiliation[c]{The Raymond and Beverly Sackler School of Physics and Astronomy,
Tel Aviv University, Ramat Aviv 69978, Israel}
\emailAdd{shimonya@tauex.tau.ac.il}
\abstract{
We classify the non-Abelian anomaly of the Euclidean conformal group $SO(2n+1,1)$ in $2n$ dimensions via Stora-Zumino descent from its Euler invariant polynomial in $2n+2$ dimensions. In this way, we place the conformal anomaly on the same footing as ordinary perturbative ’t Hooft anomalies. We also explore the relation of the non-Abelian anomaly to the known \textit{type-A Weyl anomaly}, which involves projecting into a Weyl cocycle. We discuss implications for anomaly inflow, and 't Hooft anomaly matching for the full conformal group with a Wess-Zumino-Witten term. In 4d, this enables the construction of a dilaton effective action matching the full non-Abelian $SO(5,1)$ conformal anomaly.
}
\maketitle

\section{Introduction}
Perturbative anomalies are obstructions to the gauging of global symmetries. When a global symmetry has a 't Hooft anomaly, we cannot couple its currents to a background field in a gauge invariant way; background gauge transformations lead to a nontrivial phase of the path integral, originating from the variation of the path-integral measure. This variation is cohomologically non-trivial, namely, it cannot be removed by local counterterms. The cohomological classification of perturbative chiral anomalies was done a long time ago by Wess and Zumino (WZ) \cite{Wess:1971yu}, who wrote down a consistency condition for all perturbative anomalies. They then showed that nearly all known perturbative anomalies in $2n$ dimensions arise via Stora-Zumino (SZ) descent from an invariant polynomial -- a top form in $2n+2$ dimensions. For example the famous $SU(N)$ anomaly in 4D arises from the 6D \textit{Chern character} in 6D, $I^{\rm Chern}_6={\rm tr}\left[F\wedge F \wedge F\right]$,  via Stora-Zumino (SZ) descent \cite{Zumino:1983ew,Stora:1983ct,Manes:1985df}.

The fact that most perturbative anomalies arise by descent allows us to further characterize them via \textit{Anomaly Inflow} \cite{CallanHarvey1985}. This is done in several steps: (a) turning on background gauge fields for all symmetries participating in the anomaly of the $2n$ dimensional theory.
The anomaly in this case leads to a nontrivial multiplicative factor\footnote{This factor is a phase in the case of chiral anomalies, and a rescaling factor for conformal anomalies.} of the path integral upon a gauge transformation of the background field; (b)
considering the $2n$ dimensional theory as the boundary of a $2n+1$ dimensional bulk carrying a particular Chern-Simons (CS) theory for the background gauge fields, a.k.a the \textit{anomaly theory} \cite{Kapustin2014SPTCohomology,Witten2016FermionPathIntegral}. The bulk CS theory is chosen so that its variation under a background gauge transformation localizes to the boundary and exactly cancels against the anomaly of the boundary theory. SZ Descent provides an immediate recipe for finding the appropriate CS theory: it is the $2n+1$-form $Q_{2n+1}$ whose exterior derivative is the original invariant polynomial from which the anomaly descends, $I_{2n+2}=dQ_{2n+1}$. Given a particular anomaly polynomial $I_{2n+2}$ for the anomaly, $Q_{2n+1}$ can immediately be found via the CS formula \cite{Chern:1974ft}. This guarantees that the variation of the CS term cancels against the anomaly, up to local (cohomologically trivial) terms that can be removed by adding a local counterterm of background gauge fields on the boundary.

The existence of an anomaly theory for most perturbative anomalies allows us to generalize the notion of anomalies to any theory participating in anomaly inflow with a higher dimensional bulk \cite{CallanHarvey1985,Kapustin2014SPTCohomology,FreedHopkins2016}. This larger framework encompasses far more than vanilla perturbative anomalies, tying them together with
global anomalies \cite{CallanHarvey1985,Witten1989,Witten2016FermionPathIntegral},
discrete anomalies \cite{DijkgraafWitten1990,Csaki:1997aw,KapustinThorngren2014,KapustinSeiberg2014,HsinLam2019},
anomalies of higher symmetries \cite{GaiottoKapustinSeibergWillett2015,ThorngrenElse2018,FreedHopkins2016,Cheng:2025ube},
and anomalies of non-invertible symmetries \cite{FuchsSchellekensSchweigert1996,BaisSlingerland2009,AasenMongFendley2020,
CordovaOhmori2019,BeniniTachikawa2022},
among others.

There is, however, one important type of anomalies that so far largely escaped characterization via anomaly inflow: the \textit{conformal anomalies}. These are the anomalies of a classical Weyl-invariant theory under Weyl (local scale) transformations \cite{Duff:1977ay,Bonora:1985cq,Deser:1976yx,Duff:1993wm,Deser:1993yx,Schwimmer:2010za}. Topologically non-trivial conformal anomalies are referred to as \textit{type-A} conformal anomalies, and they have famously been used to explore the non-invertibility of renormalization group flow in 2d \cite{Zamolodchikov:1986gt}, 4D \cite{Cardy1988,Komargodski:2011vj,Komargodski:2011xv}, and 6D \cite{Casini:2012ei}. Cohomological classifications of these anomalies were carried out in  \cite{Bonora:1985cq,Deser:1993yx,Boulanger:2007ab,Rovere:2024nwc} (see also \cite{Karakhanian:1994yd} for a classification in terms of diff x Weyl). 

A related, but not identical anomaly is the non-Abelian anomaly of the full $SO(2n+1,1)$ Euclidean conformal group, which includes dilatations as a subgroup. Note, however that the Weyl transformations widely considered in the literature (namely rescalings of a Riemannian metric) are not exactly dilatations; instead, they are dilatations followed by a vielbein-dependent compensating special conformal transformation imposing \textit{Cartan constraints} \cite{Kaku:1977pa}. In other words, Weyl transformation are not linear but rather \textit{non-linear} $SO(5,1)$ transformations, and their cohomology differs from that of the $SO(5,1)$ non-Abelian anomaly studied in this paper. Nevertheless, we will show how to reproduce the type-A Weyl anomaly as the projection of the full $SO(5,1)$ into a Weyl cocycle.

As we will show in this work, non-Abelian conformal anomalies in fact do have an anomaly theory; the latter already appeared in \cite{Apruzzi:2025hvs} in the context of the symmetry topological field theory (symTFT) \cite{KaidiNardoniZafrirZheng2023,Bhardwaj:2024igy,Bonetti:2024cjk} for the spontaneous breaking of conformal symmetry to Poincar\'{e} \cite{Rabinovici:2007hz}. An alternative route was pursued in \cite{Imbimbo:2023sph,Imbimbo:2025ffw} where the (super-)conformal anomaly were classified via generalized descent  with  generalized gauge fields involving BRST ghost number. In another recent approach \cite{Bittleston:Costello:2510.26764}, conformal anomalies are computed from a holomorphic formulation in 5d \textit{twistor space}. More traditionally, Weyl anomalies have a long-standing history of holographic computations in $AdS$ space \cite{Henningson:1998gx,Henningson:1998ey,Imbimbo:1999bj,Schwimmer:2003eq,Banados:2004zt,Nicolis:2008in,Coradeschi:2013gda}. Here we take the complementary step of identifying the ``classic" SZ Descent relevant to the non-Abelian conformal anomaly in $2n$ dimensions; in fact, we show that these anomalies simply descend from the $2n+2$ \textit{Euler Class} for the gauge group $SO(2n+1,1)$ identified either as the $2n$ dimensional Euclidean conformal group or the $2n+2$ dimensional Lorentz group. In this way, we can position these anomalies on par with all other perturbative anomalies. We briefly discuss their relation to the familiar expressions \cite{Deser:1993yx} for the type-A \text{Weyl} anomaly upon specializing to a \textit{Riemannian} background, which is subject to the usual Cartan constraints on the background fields of $SO(5,1)$ -- though, as we clarified above, Weyl and pure dilatations are related, but not identical. As expected, the non-Abelian conformal anomaly is both \textit{real} (not a phase) and \textit{not quantized}, 
as the Euler invariant polynomial for $SO(5,1)$ does not define a genuine (integer valued) characteristic class, namely $H^6(BSO(5,1),\mathbb{Z})$. In other words, there are no $SO(5,1)$ ``6d instantons". 

Our descent derivation further allows us to write down the corresponding \textit{dilaton effective action}
\cite{Fubini:1976jm,Deser:1993yx,Schwimmer:2010za,Komargodski:2011vj,Coradeschi:2013gda}. In our case, we can write down a dilaton effective action realizing the full conformal/Poincar\'{e} symmetry, by starting from the Wess-Zumino-Witten (WZW) \cite{Wess:1971yu,Witten:1983ar} term for $SO(5,1)/ISO(4)$, already explored in \cite{Apruzzi:2025hvs}, and substituting the inverse-Higgs relation \cite{IvanovOgievetsky1975}. We leave the analysis of scattering in this effective action, and its potential implications for RG flows, for future work.

Textbook discussions of conformal symmetry typically define conformal currents in flat space using the (classically traceless) energy–momentum tensor together with conformal Killing vectors. As a result, the conformal currents are not independent, in contrast with the currents associated with an ordinary non-Abelian internal symmetry. This feature persists on general Riemannian backgrounds \footnote{We are grateful to Zohar Komargodski for stressing this important point.}.

From the background-field perspective, restricting to Riemannian geometry amounts to considering only a subset of $SO(2n+1,1)$ background gauge fields, namely those satisfying the Cartan constraints of vanishing dilatation gauge field and torsion. One might therefore wonder whether, under this restriction, anomaly matching conditions should be correspondingly weakened, so that only the anomaly on the constrained (Riemannian) background space must be matched.

In this work we adopt a different standpoint. We treat conformal symmetry as a genuine non-Abelian global symmetry $SO(2n+1,1)$, and we demand ’t Hooft anomaly matching for the full symmetry, without imposing any constraints on the background fields or on the associated currents. This is in the original spirit of anomaly matching: a theory with a global non-Abelian symmetry should remain consistent upon coupling to arbitrary background gauge fields for all generators of the symmetry. In the present case, this requires matching the conformal anomaly even in backgrounds with torsion or a non-vanishing dilatation gauge field, where neither the background fields nor the currents satisfy the Cartan constraints. Anomaly matching is therefore a consistency condition on the path integral under coupling to arbitrary background gauge fields for the symmetry, independent of any geometric interpretation of those backgrounds.

The paper is organized as follows. In Sec.~\ref{sec:SZdescent} we review the standard construction of the Stora-Zumino descent for chiral anomalies. In Sec.~\ref{sec:desc} we present our proposal for using the Euler invariant polynomial for $SO(2n+1,1)$ as the starting point for the descent into the $2n$ dimensional conformal anomaly, and present an explicit 2 dimensional check. In Sec.~\ref{sec:4dEu} we apply our approach to classify the non-Abelian conformal anomaly in 4d. In Sec.~\ref{sec:dilaton} we derive a 4d dilaton effective action corresponding to conformal/Poncar\'{e} spontaneous symmetry breaking (SSB); this action naturally arises in IR EFTs to match the full $SO(5,1)$ anomaly of the UV. In Sec.~\ref{sec:2ndims} we generalize our results to $2n$ dimensions, and present the general anomaly inflow. We conclude in Sec.~\ref{sec:conclusions}, where we also present some potential future directions. In Appendix~\ref{app:SU2} we connect our 2d Euler/Pontryagin approach to the prevalent derivation of 2d conformal and gravitational anomalies as the sum and difference of left and right movers. Finally in Appendix~\ref{app:SU4} we reformulate the $SO(5,1)$ anomaly in 4d in the equivalent language of, $SU^*(4)$ the noncompact real form of $SU(4)$. 

\section{Stora-Zumino descent\label{sec:SZdescent}}
The WZ consistency condition on perturbative anomalies for a Lie group $G$ comes from the definition (see, e.g. \cite{Bilal:2008qx}):
\begin{eqnarray}
\mathcal{A}_{\hat{a}}(x)=-\left(D_\mu \frac{\delta}{\delta A_\mu(x)}\right)_{\hat{a}} \Gamma[A],
\end{eqnarray}
where $\mathcal{A}_{\hat{a}}$ is the anomaly, $\Gamma[A]$ is the effective action for the background gauge field $A_\mu$ for $G$, and $D_\mu$ is the gauge-covariant derivative. The fact that the anomaly is a \textit{covariant derivative} of a functional derivative (in this case of the effective action), immediately constrains the possible expression for $\mathcal{A}_a$ via
\begin{eqnarray}\label{eq:WZcon}
\left(D_\mu \frac{\delta}{\delta A_\mu(x)}\right)_{\hat{a}} \mathcal{A}_{\hat{b}}-\left(D_\mu \frac{\delta}{\delta A_\mu(y)}\right)_{\hat{b}} \mathcal{A}_{\hat{a}}(x)=-f_{\hat{a}\hat{b}\hat{c}}\,\delta(x-y)\,\mathcal{A}_{\hat{c}}(x)\,,
\end{eqnarray}
where the $f_{abc}$ are the structure constants of the group $G$. This is the WZ consistency condition (see e.g. \cite{Bilal:2008qx} for the proof). For $G=SU(N)$, this condition is solved (up to local terms) by SZ Descent; to see this, start from the Chern character in $2n+2$ dimensions:
\begin{eqnarray}
I^{\rm Chern}_{2n+2}=\frac{1}{(n+1)!}\left(\frac{i}{2 \pi}\right)^{n+1} {\rm tr}\left(F^{n+1}\right)\,,
\end{eqnarray}
where $F=dA+A\wedge A$, ${\rm tr}$ is the $SU(N)$ trace, and $F^{n+1}=F\wedge\ldots\wedge F$, $n+1$ times. A closed form expression for the \textit{Chern-Simons form} $Q^{\rm Chern}_{2n+1}$ satisfying $dQ^{\rm Chern}_{2n+1}=I^{\rm Chern}_{2n+2}$ is
\begin{eqnarray}
Q^{\rm Chern}_{2n+1}&=&\frac{1}{n!}\left(\frac{i}{2 \pi}\right)^{n+1}\,{\rm tr} \left[ \int_0^1 d t \,A\wedge F^{n}_t\right]\,\nonumber\\[5pt]
F_t&\equiv& t dA+t^2 A\wedge A\,.
\end{eqnarray}
where again the $n^{th}$ power indicates an $n$-fold wedge product. For $n=1$ and $n=2$, respectively:
\begin{eqnarray}
Q^{\rm Chern}_{3}&=&\frac{1}{2}\left(\frac{i}{2\pi}\right)^2\,{\rm tr} \left(A\wedge dA+\frac{2}{3}A^3\right)\,\nonumber\\[5pt]
Q^{\rm Chern}_{5}&=&\frac{1}{3!}\left(\frac{i}{2\pi}\right)^3\,{\rm tr}\left(A\wedge dA^2+\frac{3}{2}\,A^3\wedge dA+\frac{3}{5} A^5\right)\,.
\end{eqnarray}
Under an infinitesimal transformation $\varepsilon\equiv \varepsilon_{\hat{a}}T^{\hat{a}}$, the gauge field transforms as $\delta_\varepsilon A = D\varepsilon=d\varepsilon+[A,\varepsilon]$. Under this transformation, the variation of the CS form $Q^{\rm Chern}_{2n+1}$ is 
\begin{eqnarray}
\delta_\varepsilon Q^{\rm Chern}_{2n+1}&=&dQ^{\rm Chern,1}_{2n}(\varepsilon,A)\,,
\end{eqnarray}
where
\begin{eqnarray}
Q^{\rm Chern,1}_{2n}(\varepsilon,A)=\frac{1}{(n-1)!}\left(\frac{i}{2 \pi}\right)^{n+1}\,{\rm tr} \left[\varepsilon \, \int_0^1 d t\,(1-t)\, \,d\left(A\wedge F^{n-1}_t\right)\right]\,.
\end{eqnarray}
The crux is that the anomaly defined by
\begin{eqnarray}
\int d^{2n}x\,\varepsilon_{\hat{a}} \mathcal{A}^{\hat{a}}=\int d^{2n}x\,Q^{\rm Chern,1}_{2n}(\varepsilon)\,,
\end{eqnarray}
automatically satisfies the WZ consistency conditions \cite{Wess:1971yu}. Finally we note that we can always add a total derivative $d\alpha_{2n}$ to $Q^{\rm Chern}_{2n+1}$ while keeping $I^{\rm Chern}_{2n+2}=dQ^{\rm Chern}_{2n+1}$, or add a total derivative $d\beta_{2n-1}$ to $Q^{{\rm Chern},1}_{2n}$ without changing $\delta_\varepsilon Q^{\rm Chern}_{2n+1}=dQ^{{\rm Chern},1}_{2n}$. This means that the anomaly $Q^{{\rm Chern},1}_{2n}$ is only defined up to 
\begin{eqnarray}
Q^{{\rm Chern},1}_{2n}\sim Q^{{\rm Chern},1}_{2n}+\delta_\varepsilon \alpha_{2n}+d\beta_{2n-1}\,.
\end{eqnarray}
These shifts move us between different representatives in the cohomology class of the anomaly. Note that for $SU(N)$, the descent procedure gives the unique solution to the WZ consistency conditions, modulo local terms \cite{Barnich:2000zw}. This would not necessarily be the case below when we work out the descent for $SO(2n+1,1)$.

\section{Descent for $SO(2n+1,1)$ in $2n$ dimensions}\label{sec:desc}
Consider an $SO(2n+1,1)$ gauge bundle. We wish to classify its anomalies in $2n$ dimensions via descent from $2n+2$ dimensions. We will label $SO(2n+1,1)$ generators via the antisymmetric vector labeling $T^{ij}$ rather than the adjoint index $a$, and work in the convention where these generators are \textit{anti-hermitian}. Equivalently, we can form linear combinations of the $T^{ij}$ generators making up the generators of the $2n$ dimensional Euclidean conformal group\footnote{In 2d this is just its ``global part".} \cite{KakuTownsendVanNieuwenhuizen1977,KakuTownsendVanNieuwenhuizen1978,KugoUehara1985}:
\begin{eqnarray}\label{eq:gens}
D=T^{2n,2n+1}~,~M^{ab}=T^{ab}~,~P^a=T^{a,2n+1}-T^{a,2n}~,~K^a=T^{a,2n+1}+T^{a,2n}\,.
\end{eqnarray}
Where $D,\,M^{ab},\,P^a,\,K^a,\,a=\{0,...,2n-1\}$ are the generators for dilatations, Lorentz transformations, translations, and special conformal transformations, respectively. We can turn on background gauge fields for these generators as \cite{DiFrancescoMathieuSenechal1997}
\begin{eqnarray}\label{eq:gensgauge}
\frac{1}{2}A^{ij} T_{ij}=b D + \frac{1}{2}w^{ab} M^{ab}+ e^{a} P^{a}+ f^{a} K^{a}\,.
\end{eqnarray}
Where there is no distinction between upper and lower indices $a,b$.
Interestingly, there are two different invariant polynomial for $SO(2n+1,1)$ in $2n+2$ dimensions (see e.g. \cite{KobayashiNomizu1963,MilnorStasheff1974,Nakahara:2003nw}): \begin{itemize}
\item The Euler invariant polynomial
\begin{eqnarray}\label{eq:gen Euler}
I^{\rm Euler}_{2n+2}={\rm Pf}\left(\frac{F}{2 \pi}\right)=\frac{\epsilon_{i_1 \cdots i_{2 n+2}}}{(n+1)!(4 \pi)^{n+1}}  F^{i_1 i_2} \wedge \cdots \wedge F^{i_{2 n+1} i_{2 n+2}} .
\end{eqnarray}
 This invariant polynomial is even under CP, and so is the natural candidate to descend into the (CP-even) non-abelian conformal anomaly.
\item Pontryagin invariant polynomial. This is a real invariant polynomial denoted by $p_{k}$. The first few Pontryagin invariant polynomials are
\begin{eqnarray}
p_1&=&-\frac{1}{8\pi^2}{\rm tr}\left(F^2\right)\nonumber\\[5pt]
p_2&=&\frac{1}{128\pi^4}\left[{\rm tr}\left(F^2\right)^2-2{\rm tr}\left(F^4\right)\right]\nonumber\\[5pt]
p_3&=&-\frac{1}{3072\pi^6}\left[{\rm tr}\left(F^2\right)^3-6{\rm tr}\left(F^2\right){\rm tr}\left(F^4\right)+8{\rm tr}\left(F^6\right)\right]\,.
\end{eqnarray}
When $n+1=2m$, we can find a product of Pontryagin invariant polynomials that make up a $2n+2$-form. In that case, consider $k_1+\ldots+k_r=m$ for some $r>0$. Then $p_{k_1}\wedge\ldots\wedge p_{k_r}$ is a $2n+2$ dimensional invariant polynomial.
In particular, $I^{\rm Pontryagin}_{2n+2}=p_{m}$ for $n+1=2m$ is a $2n+2$ dimensional invariant polynomial. It is odd under CP, and, indeed, it is known to descent into the (CP odd) gravitational anomaly in $4m-2$ dimensions \cite{AlvarezGaumeWitten1984,AlvarezGaumeGinsparg1985}.
\end{itemize}

In this paper we will consider descent from both $I^{\rm Euler}_{2n+2}$ and $I^{\rm Pontryagin}_{2n+2}$ for $SO(2n+1,1)$, which we identify as the $2n$ dimensional conformal group (or just its global part for $n=1$). As we will see explicitly, the anomaly descending from $I^{\rm Euler}_{2n+2}$ is the non-Abelian conformal anomaly in $2n$ dimensions. This is in contrast with the anomaly descending from $I^{\rm Pontryagin}_{2n+2}$, which is famously (see e.g. \cite{Bertlmann:1996xk}) the gravitational anomaly. A similar-in-spirit discussion of $2n+1=3$ dimensional CS forms appears in theories of 3d gravity \cite{AchucarroTownsend1986,Witten1988,Banados:2004zt,Witten:2007kt}. Note that unlike the $SU(N)$ case \cite{Barnich:2000zw}, we do not show uniqueness of the descent cohomology as a solution for the WZ consistency conditions. This leaves the possibility of other solutions to the WZ consistency conditions, for example the type-B anomaly. We do not address this anomaly in this manuscript.

We begin with low dimensional examples for simplicity. For $n=1$ we have
\begin{eqnarray}
I^{\rm Euler}_{4}=\frac{1}{32 \pi^{2}}  \epsilon_{i_1 i_2 i_3 i_{4}}F^{i_1 i_2} \wedge F^{i_{3} i_{4}} .
\end{eqnarray}
The corresponding Chern-Simons form is
\begin{eqnarray}
Q^{\rm Euler}_{3}=\frac{1}{32 \pi^{2}}   \epsilon_{i_1 i_2 i_3 i_{4}}\left(A^{i_1i_2}\wedge dA^{i_3i_4}+\frac{2}{3}A^{i_1i_2}\wedge A^{i_3j}\wedge A_{j}{}^{i_4}\right)\, .
\end{eqnarray}
This form appears in formulations of 3d gravity \cite{AchucarroTownsend1986,Witten1988,Witten:2007kt}, and recently as part of the SymTFT construction for Conformal/Poincar\'{e} spontaneous symmetry breaking (SSB) in 2d \cite{Apruzzi:2025hvs}. Finally, an infinitesimal $SO(3,1)$ transformation gives
\begin{eqnarray}\label{eq:2danomaly}
Q^{\rm Euler,\,1}_{2}(\varepsilon,A)&=&\frac{1}{32 \pi^{2}}  \epsilon_{i_1 i_2 i_3 i_4}\varepsilon^{i_1,i_2}dA^{i_3i_4}\,.
\end{eqnarray}
This is our general result for the 2d conformal anomaly.
\subsection{Reproducing the 2d Weyl Anomaly}\label{sec:2dWeyl}
The 2d Weyl anomaly is an anomaly under local rescalings of a Riemannian metric. Namely, given a Riemannian metric $g_{\mu\nu}$, an infinitesimal \textit{Weyl transformation} with parameter $\sigma(x)$ is given by 
\begin{eqnarray}\label{eq:2dWeyl}
\delta^{\rm Weyl}_\sigma g_{\mu\nu}=-2\sigma g_{\mu\nu}\,.
\end{eqnarray}
To reproduce the Weyl anomaly from our general expression \eqref{eq:2danomaly}, we first specialize to background $e^a,\,f^a,\,w^{ab}$ and $b$ equivalent to a Riemannian metric. While $e^a$ is generic, $w^{ab}$ and $b$ are fixed by Cartan constraints to be
\begin{eqnarray}\label{eq:cartan}
b=0~~,~~w^{ab}=\omega^{ab}\,,
\end{eqnarray}
where $\omega^{ab}$ is the torsionless spin connection of $e^a$ satisfying
\begin{eqnarray}\label{eq:contor}
&&de^{a}+\,\omega^{a}_{~~b}\wedge e^{b}=0\,.
\end{eqnarray}
In a number of dimensions $n>2$, $f^a$ is also fixed by the Cartan constraints, while in 2d it is not. We will return to this point when we generalize our classification to higher dimensions. In 2d, an infinitesimal Weyl transformation \eqref{eq:2dWeyl} acts on the background fields as 
\begin{eqnarray}\label{eq:dWeyl}
\delta^{\rm Weyl}_\sigma b=0~~,~~\delta^{\rm Weyl}_\sigma w^{ab}=\nabla^{[a}\sigma e^{b]}\,,
\end{eqnarray}
where $\nabla^a \sigma=e^{a\mu}\partial_\mu\sigma$. In 2d, the transformation on $f^a$ is not determined by the Cartan constraints, so we can choose a convenient one,
\begin{eqnarray}\label{eq:df2d}
\delta^{\rm Weyl}_\sigma f^a=-\frac{1}{2}D(\nabla^{a}\sigma)+\sigma f^a\,~~({\rm choice\,in}\,2d)
\end{eqnarray}
and $Dv^a=dv^a+w^a_c\wedge v^c$. For this choice for the action of a Weyl transformation on $f^a$, it completely coincides with an $SO(3,1)$ transformation with parameter $\varepsilon_{\rm Weyl}=\sigma D-\frac{1}{2}\nabla^a \sigma K^a$. Substituting it in the general expression \eqref{eq:2danomaly}, we get 
\begin{eqnarray}\label{eq:2dA}
Q^{\rm Euler,\,1}_{2}(\varepsilon_{\rm Weyl},A)|_{\rm Riem}+\delta_{\varepsilon_{\rm Weyl}} \alpha_2+d(\ldots)=\frac{\sigma }{16 \pi^{2}}  d\omega=\frac{\sigma }{16 \pi^{2}}  \sqrt{g}Rd^2x\,,
\end{eqnarray}
where $\alpha_2$ is a local counterterm,
\begin{eqnarray}\label{eq:alpha2beta1}
\alpha_2=-\frac{1}{4\pi^2}\epsilon_{ab} f^a\wedge e^b\,.
\end{eqnarray}

The 2d Weyl anomaly is then
\begin{eqnarray}\label{eq:res2d}
\delta^{\rm Weyl}_\sigma \Gamma=8\pi c \int \,\left[Q^{\rm Euler,\,1}_{2}(\varepsilon_{\rm Weyl})|_{\rm Riem}+\delta_{\varepsilon_{\rm Weyl}}\alpha_2\right]=c\int \frac{\sigma }{4\pi}  \sqrt{g}Rd^2x\, .
\end{eqnarray}
We normalized it\footnote{We choose this normalization for consistency with the $2n$ dimensional result. It is related to the standard definition $c_{standard}$ as $c_{standard}=6c$.} so that $\delta_\sigma \Gamma=2c$ on $S^2$ for $\sigma=1$. We have thus reproduced the 2d Weyl anomaly up to local terms via SZ descent from the 4d Euler invariant polynomial for $SO(3,1)$. We wish to emphasize that the general result for the 2d conformal anomaly is in fact \eqref{eq:2danomaly}. The known result \eqref{eq:res2d} is a special case of this more general result, when the background is Riemannian, and when the infinitesimal variation coincides with an infinitesimal Weyl transformation. Furthermore, it would be interesting to realize the more general case by an explicit construction of a classically conformally invariant theory in a torsionful background. We revisit this point in the conclusions.

Here is the place to comment on an important order-of-operations subtlety; one may wonder whether we should also transform $\omega$ under Weyl transformations in a way that invalidates the derivation of \eqref{eq:res2d}. This is not the case. In fact, the Cartan constraints \eqref{eq:cartan} do not commute with gauge transformations, and we only apply it \textit{after} doing a background gauge transformation and deriving the anomaly. Reversing the order of operations would not make sense, as then we would have to either (a) gauge transform out of the constraint, or (b) combine a background gauge transformation with a compensating background non-gauge transformation to preserve the constraint. In either case we would not have correctly captured the anomaly.
\subsection{Aside: reproducing the known 2d Gravitational anomaly}

The analogous exercise starting from the 4d Pontryagin invariant polynomial gives
\begin{eqnarray}
I^{\rm Pontryagin}_{4}&=&-\frac{1}{8\pi^{2}}  {\rm tr}\left(F\wedge F\right)\nonumber\\[5pt]
Q^{\rm  Pontryagin}_{3}&=&-\frac{1}{8\pi^{2}}   {\rm tr}\left(A\wedge dA+\frac{2}{3}A^3\right)\nonumber\\[5pt]
Q^{\rm  Pontryagin,\,1}_{2}(\varepsilon)&=&-\frac{1}{8\pi^{2}}  {\rm tr}\left(\varepsilon dA\right)\, .
\end{eqnarray}
Here $Q^{\rm  Pontryagin}_{3}$ is equivalent to the gravitational CS term used in \cite{Chamseddine:1992ry} for the 2d gravitational anomaly. Selecting $\varepsilon=\lambda M^{01}$ and applying $b=0$ and the vanishing torsion condition, we get
\begin{eqnarray}\label{eq:res2dpont}
\delta_\lambda \Gamma=-\int 4\pi c_{grav} \,Q^{\rm Pontryagin,\,1}_{2}({\rm Lorentz})=c_{grav}\int \frac{\lambda }{4\pi}  \sqrt{g}Rd^2x\, .
\end{eqnarray}
We thus reproduce the gravitational anomaly \cite{AlvarezGaumeWitten1984,AlvarezGaumeGinsparg1985} by descending from the 4d Pontryagin invariant polynomial. Since $SO(3,1)$ is isomorphic to the noncompact form of $SU(2)_L\times SU(2)_R$, we can easily present the ``Euler" and ``Pontryagin" anomalies as linear combinations of the left and right contributions to the anomaly -- see Appendix~\ref{app:SU2} for the details. In this way we reproduce the well known fact \cite{AlvarezGaumeWitten1984,AlvarezGaumeGinsparg1985,Chamseddine:1992ry} that the gravitational and Weyl anomalies correspond to the difference and sum of the left mover and right mover, respectively.
\section{The 4d non-Abelian conformal anomaly from Euler descent}\label{sec:4dEu}
The non-Abelian conformal anomaly, and its cousin the type-A Weyl anomaly, are of utmost importance to RG flows in 2d \cite{Zamolodchikov:1986gt} and 4d \cite{Komargodski:2011vj,Komargodski:2011xv}. Here we obtain the 4d non-Abelian conformal anomaly by descent from the 6d Euler invariant polynomial for $SO(5,1)$, the 4d Euclidean conformal group. For ease of reading, we introduce the notation
\begin{eqnarray}
\etr{X}{Y}{Z}\equiv\epsilon_{i_1 i_2 i_3 i_4 i_5 i_6}X^{i_1i_2}\wedge Y^{i_3i_4}\wedge Z^{i_5i_6}\,,
\end{eqnarray}
as well as $[XY]^{ij}\equiv X^{i}_{~k}Y^{kj}$. We begin with the Euler invariant polynomial for $SO(5,1)$ in 6d.
\begin{eqnarray}
I^{\rm Euler}_{6}=\frac{1}{384 \pi^{3}}  \etr{F}{F}{F}\, .
\end{eqnarray}
The CS form $Q^{\rm Euler}_5$ satisfying $I^{\rm Euler}_{6}=dQ^{\rm Euler}_5$ is
\begin{eqnarray}\label{eq:Q5Euler}
Q^{\rm Euler}_{5}=\frac{1}{384  \pi^{3}}  &&\left[\etr{A}{dA}{dA}+\frac{3}{2}\etr{A}{A^2}{dA}+\frac{3}{5} \etr{A}{A^2}{A^2}\right]\,.
\end{eqnarray}
This is the same CS form used in \cite{Apruzzi:2025hvs} as part of the SymTFT for Conformal/Poincar\'{e} SSB in 4 dimensions. Varying the CS form with $\varepsilon=\varepsilon_{i_1i_2}T^{i_1i_2}$ we have $\delta_\varepsilon Q^{\rm Euler}_{5}=dQ^{\rm Euler,1}_{4}(\varepsilon)$ with
\begin{eqnarray}\label{eq:4dan}
Q^{\rm Euler,1}_{4}(\varepsilon,A)&=&\frac{1}{384 \pi^{3}}  \epsilon_{i_1 i_2 i_3 i_4 i_5 i_6}\varepsilon^{i_1 i_2} d\left(A^{i_3 i_4}\wedge dA^{i_5 i_6}+\frac{1}{2}A^{i_3 i_4}\wedge A^{i_5 j}\wedge A_{j}^{~i_6}\right)\,.
\end{eqnarray}
The expression \eqref{eq:4dan} is the general result for the 4d non-Abelian conformal anomaly. Since $SO(5,1)$ is isomorphic to $SU^*(4)$, the non-compact form of $SU(4)$, we could have derived the same anomaly by descending from the 3rd Chern character of the latter\footnote{We thank Adam Schwimmer for his question regarding the potential equivalent formulation of our $SO(5,1)$ anomaly in terms of $SU^*(4)$.}. This route, and its equivalence to the present derivation, is shown in Appendix~\ref{app:SU4}. Finally, we can relate \eqref{eq:4dan} to an expression made only out of the Lorentz connection $w^{ab}$ using CS transgression,
\begin{eqnarray}\label{eq:transgression}
Q^{{\rm Euler},1}_4(\varepsilon,A)=Q^{{\rm Euler},1}_4(\varepsilon,A_w)+\delta_\varepsilon B_4+d(\ldots)\,,
\end{eqnarray}
where $A_w$ is $A$ with all components but $w^{ab}$ set to zero. The local counterterm $B_4$ is given by \cite{Nakahara:2003nw}
\begin{eqnarray}\label{eq:B4}
B_4=\frac{6}{384\pi^2}\int_0^1dt\int_0^1 s\,\left\langle A_t,A-A_w,F_{st}\right\rangle\,,
\end{eqnarray}
and $F_{st}=dA_{st}+A_{st}\wedge A_{st}$, $A_{st}=s[A_w+t(A-A_w)]$.

\subsection{Comparison with type-A Weyl anomaly}
As we noted in Section~\ref{sec:2dWeyl}, Weyl transformations are local rescaling of the metric/vielbein that respect the Cartan constraints. While in 2d we had the freedom of completing the transformation for $f^a$ so that Weyl transformation coincide with infinitesimal conformal transformations, we do not have this luxury in 4d. In $n>2$ dimensions, the Cartan constraints are
\begin{eqnarray}\label{eq:cartan4d}
b=0~~,~~w^{ab}=\omega^{ab}~~,~~f^a=-\frac{1}{2}P^a_{~b}e^b\,,
\end{eqnarray}
where $P^{a}_{~b}$ is the Schouten tensor
\begin{eqnarray}\label{eq:contor}
P_{ab}\,e^a_{\mu}e^b_{\nu}=\frac{1}{n-2}\left(R_{\mu\nu}-\frac{1}{2(n-1)}Rg_{\mu\nu}\right)\,,
\end{eqnarray}
and $g_{\mu\nu}=\eta_{ab}e^a_\mu e^b_\nu$. We take the number of dimensions to be $n=4$ in this section. This means that infinitesimal Weyl transformations in 4d are
\begin{eqnarray}\label{eq:dWeyl4}
\delta^{\rm Weyl}_\sigma b=0~~,~~\delta^{\rm Weyl}_\sigma w^{ab}=\nabla^{[a}\sigma e^{b]}~~,~~\delta^{\rm Weyl}_\sigma f^a=-\frac{1}{2}D(\nabla^{a}\sigma)+\sigma f^a\,,
\end{eqnarray}
and they respect the Cartan constraints \eqref{eq:cartan} by definition. Once again, this infinitesimal Weyl transformation coincides with an infinitesimal $SO(5,1)$ transformation in the $\varepsilon=\varepsilon_{\rm Weyl}\equiv\sigma D-\frac{1}{2}\nabla^a \sigma K^a$. However, note that the $\nabla^a$ in $\varepsilon_{\rm Weyl}$ \textit{depends on the vielbein} $e^a$ as $\nabla^a=e^{a\mu}\partial_\mu$. In other words, Weyl transformations are not ordinary gauge transformations but \textit{nonlinear ones}.

A direct outcome of the nonlinearity of Weyl transformations on $e,\,\omega,\,f$ is that our general result for the non-Abelian anomaly \eqref{eq:4dan}, while satisfying the Wess-Zumino consistency condition \eqref{eq:WZcon} for $SO(5,1)$, does not in fact reproduce the condition for \textit{Weyl transformations}. Indeed, the non-linearity of the latter makes the consecutive application of two $\delta_{\varepsilon_{\rm Weyl}(e)}$ behave differently from the case of two \textit{vielbein independent} transformations $\delta_{\varepsilon_{SO(5,1)}}$. To mend this, we have to project \eqref{eq:4dan} into a Weyl cocycle.

To see this, we first specialize \eqref{eq:transgression} to $\varepsilon=\varepsilon_{\rm Weyl}$ and $w=\omega$:
\begin{eqnarray}\label{eq:transgression2}
&&Q^{{\rm Euler},1}_4(\varepsilon_{\rm Weyl},A_{\omega})= \frac{\sigma}{384 \pi^{3}}  \epsilon_{abcd}d\left[\omega^{ab}\wedge\left(d\omega^{cd}+\frac{2}{3}\omega^{c}_{~e}\wedge\omega^{ed} \right)\right]-\Delta Q^1_{4\,{\rm Weyl}}\nonumber\\[5pt]
&&\Delta Q^1_{4\,{\rm Weyl}}=\frac{\sigma}{6\times 384 \pi^{3}}  \epsilon_{abcd}d\left[\omega^{ab}\wedge\omega^{c}_{~e}\wedge\omega^{ed}\right]\,.
\end{eqnarray}
In the first term, we recognize the familiar expression for the type-A Weyl anomaly, while $\Delta Q^1_{4\,{\rm Weyl}}$ is a residual term is neither Weyl-closed nor diffeomorphism-invariant, and we need to project it out to obtain a genuine Weyl cocycle. If we remove this term by hand, we get
\begin{eqnarray}\label{eq:4dtypeA0}
Q^{{\rm Euler},1}_4(\varepsilon_{\rm Weyl},A_{\omega})-\Delta Q^{4}_{1\,{\rm Weyl}}&=&\frac{\sigma}{24\pi^{3}}   E_4(\omega)\,\nonumber\\[5pt]
&=&\frac{\sigma }{192 \pi^{3}}  \sqrt{g}\left[R_{\mu \nu \rho \sigma} R^{\mu \nu \rho \sigma}-4 R_{\mu \nu} R^{\mu \nu}+R^2\right]d^4x\,.
\end{eqnarray}

The need to project out $\Delta Q^4_{1,\,{\rm Weyl}}$ highlights the difference between the non-Abelian $SO(5,1)$ conformal anomaly and its Weyl cousin, where the latter is realized non-linearly on the spin connection. This means, in particular, that the two solve slightly different cohomology problems. For this reason, anomaly matching for the full $SO(5,1)$ is related, but not identical, to anomaly matching Weyl transformations. As a final note, in $2d$ we reproduced the Weyl anomaly exactly because the $\omega$-dependent part of the anomaly was strictly Abelian; here, on the other hand, the residual term comes from $d\omega^3$. We suspect that this is linked to the existence of both type-A and type-B anomalies in $4d$, namely that there may be two different ways to project \eqref{eq:transgression2} into a Weyl cocycle, corresponding to type-A and type-B. We are currently exploring a systematic way to perform this projection.

\section{A dilaton effective action\label{sec:dilaton}}

Inspired by the seminal paper by Komargodski and Schwimmer (KS) \cite{Komargodski:2011vj}, one can interpret four dimensional renormalization group flows via spontaneous breaking of conformal symmetry. In the literature, one uses anomaly matching for the type-A \textit{Weyl anomaly}. In this case, one considers the RG flow between a UV CFT whose type-A Weyl anomaly is $a_{UV}E_4(\omega)$, and an IR CFT whose matter fields contribute $a_{IR}E_4(\omega)$. 't Hooft anomaly matching is then ensured via the contribution of the ``type-A" part of the \textit{dilaton effective action}:
\begin{eqnarray}\label{eq:SA}
S^A_\tau&=&-\frac{\Delta a}{8\pi^2}\int d^4 x \sqrt{-g}\left[\tau E_4+4\left(R^{\mu \nu}-\frac{1}{2} g^{\mu \nu} R\right) \partial_\mu \tau \partial_\nu \tau-4(\partial \tau)^2 \square \tau+2(\partial \tau)^4\right]\,,
\end{eqnarray}
where $\Delta a=a_{UV}-a_{IR}$, and we rescaled the overall normalization to match our conventions. The variation of $S^A_\tau$ under Weyl transformation provides the extra $\Delta a E_4(\omega)$ required to match the type-A anomaly of the UV theory. In flat space $g_{\mu\nu}\rightarrow\eta_{\mu\nu}$, only the last two terms survive:
\begin{eqnarray}\label{eq:SAf}
S^{A,flat}_\tau&=&-\frac{\Delta a}{8\pi^2}\int d^4 x \left[-4(\partial \tau)^2 \square \tau+2(\partial \tau)^4\right]\,.
\end{eqnarray}
In \cite{Komargodski:2011vj}, KS famously showed that $\Delta a>0$ as a positivity bound from dilaton $2\rightarrow 2$ scattering in the forward direction, mediated by \eqref{eq:SAf}.

In this section, we pursue a similar line-of-reasoning, but make use of the \textit{non-Abelian conformal anomaly} instead of the type-A Weyl anomaly. Recall that the difference between the two is that a Weyl is a \textit{nonlinear} conformal transformation on $e^a,\,\omega^{ab}$ and $f^a$, whose magnitude depends on the vielbein itself. Here we follow the general philosophy of 't Hooft anomaly matching; since $SO(5,1)$ is an exact global symmetry of the UV and IR CFTs, we need to match its anomalies between the two in the same way that we match any other anomaly of a global symmetry. The anomalies of the two fixed points are
\begin{eqnarray}
\delta_\varepsilon S_{UV,IR}&=& a_{UV/IT}\int_{M_5} Q^{{\rm Euler},1}_{4}(\varepsilon)\,,
\end{eqnarray}
where $\varepsilon=\frac{1}{2}\varepsilon_{ab}T^{ab}$ is any $SO(5,1)$ transformation. The difference between the matter contributions to the UV and IR anomalies is compensated for by the variation of the full WZW \cite{Wess:1971yu,Witten:1983ar} term for $SO(5,1)/ISO(4)$, where $ISO(4)$ is the 4d Poincar\'{e} group. In this section we write an analog of the KS dilaton effective action \eqref{eq:SA} and its flat-space limit \eqref{eq:SAf} that realizes $SO(5,1)/ISO(4)$ nonlinearly. By construction, it matches the non-Abelian conformal anomaly; it does not directly match \eqref{eq:SAf}, because Weyl cocycles are not the same as $SO(5,1)$ cocycles, as we've seen above. In particular, there is no reason to expect a local counterterm relating the two, since they belong to different Wess–Zumino cohomology problems.

We can write this WZW term explicitly. First, we construct the 5d manifold $\mathcal{M}_5\equiv\mathcal{M}_4\times[0,1]$. The 5th coordinate is $x_5\in[0,1]$. The coset $SO(5,1)/ISO(4)$ is parametrized by the Goldstone matrix $U$, defined as
\begin{eqnarray}\label{eq:coset}
U\equiv e^{x_5x_aP^a}e^{x_5\xi_a(x)K^a} e^{ x_5\tau(x) D}\,.
\end{eqnarray}
This parametrization\footnote{In the original construction of \cite{Zumino:1983ew}, $U$ is given by $\exp[-x_5G(x)]$ where $G$ is a linear combination of broken generators. We can use the Baker-Campbell-Hausdorff lemma to bring \eqref{eq:coset} to the form $\exp[-\sum_n x^n_5G_n(x)]$ where a slight generalization of \cite{Zumino:1983ew} holds.} is typical of spontaneously broken \textit{spacetime} symmetries, where one is allowed to eliminate the $\xi_a$ in favor of derivatives of $\tau$, via a procedure called \textit{inverse Higgsing} \cite{IvanovOgievetsky1975,Low:2001bw,Klein:2017npd}. In short, the nontrivial commutation relation $[P^a,K^b]=2(D\delta^{ab}+M^{ab})$ means that the Goldstones $\tau$ and $\xi^a$ for $D$ and $K^a$ are not independent degrees of freedom. They can be related as
\begin{eqnarray}\label{eq:IH}
\xi^a=-\frac{1}{2}\partial^a\tau\,,
\end{eqnarray}
which solves the \textit{inverse Higgs constraint} $[U^{-1}d_4U]_D=0$ at $x_5=1$, where $d_4$ is the 4d exterior derivative. In the presence of an $SO(5,1)$ background $A_\mu$, the WZW term is then (see e.g. \cite{Wess:1971yu,Witten:1983ar,Zumino:1983ew,Wu:1984pv,Hull:1990ms,DHoker:1984ph,DHoker:1994rdl,Mora:2006ka,Tachikawa:2016xvs,Brauner:2018zwr,Lee:2020ojw} and especially \cite{Apruzzi:2025hvs} for a related St\"{u}ckelberg term),
\begin{eqnarray}\label{eq:WZW}
S_{WZW}&=&\Delta a\int_{M_5} \Omega^{\rm Euler}_{5}\left(A,U\right)\,,
\end{eqnarray}
where $\Omega^{\rm Euler}_{5}\left(A,U\right)$ is given in Appendix~\ref{app:WZW}. Note that on $\mathcal{M}_4$ we set $x_5=1$ in \eqref{eq:coset}. Under any $SO(5,1)$ transformation, the variation of the WZW term gives \cite{Zumino:1983ew}
\begin{eqnarray}\label{eq:deltaSWZW}
\delta_\varepsilon S_{WZW}&=&=\Delta a\int_{\mathcal{M}_4} Q^{{\rm Euler},1}_{4}(\varepsilon)\,.
\end{eqnarray}
ensuring 't Hooft anomaly matching. Standard transgression argument provides an almost 4d presentation of $S_{WZW}$ for any $A$, namely
\begin{eqnarray}\label{eq:WZW4d}
S_{WZW}&=&\Delta a\left\{\int_{M_5} \Omega^{\rm Euler}_{5}\left(0,U\right)+\int_{M_4} B_4(A,U)\right\}\,.
\end{eqnarray}
where $B_4(A,U)$ is given in \eqref{eq:B4}. The first term in \eqref{eq:WZW4d} is the ungauged WZW term. In the chiral Lagrangian \cite{Witten:1983ar}, different extensions of $U$ into $\mathcal{M}_5$ lead to a shift of $\int_{M_5} \Omega^{\rm Euler}_{5}\left(0,U\right)\rightarrow \int_{M_5} \Omega^{\rm Euler}_{5}\left(0,U\right)+2\pi n$. Consequently, the coefficient of the WZW term in the chiral Lagrangian is quantized. Here, since $\pi_5\left[SO(5,1)/ISO(4)\right]=0$, all extensions of $U$ give the same integral, and so the coefficient of our WZW term is not quantized, as expected.

To compute the flat space limit of $S_{WZW}$, we set $A=A_{\rm flat}\equiv dx_aP^a$. In this case the general expression \eqref{eq:WZWgen} reduces to
\begin{eqnarray}\label{eq:WZWflat}
&&\Omega^{\rm Euler}_{5}\left(A_{\rm flat},U\right)=-\frac{1}{384  \pi^{3}}  \left\{-\tfrac{1}{2} \etr{\overline{\phi}}{\overline{\phi}^2}{\overline{W}}+\etr{\overline{\phi}}{\overline{W}}{\overline{W}}\right\}\,,
\end{eqnarray}
where $\overline{\phi}$ and $\overline{W}$ are defined in \eqref{eq:defphi}. Note that the WZW action \eqref{eq:WZW} reproducing the full non-Abelian conformal anomaly does not coincide with the KS action \eqref{eq:SA} exactly because Weyl transformations are not linear $SO(5,1)$ transformation, but rather nonlinear ones depending on the vielbein. It would be very interesting to study the implications on RGE flows from demanding $SO(5,1)$ anomaly matching rather than Weyl anomaly matching. We leave it for future work.

\section{Non-Abelian conformal anomaly in $2n$ dimensions, and anomaly inflow\label{sec:2ndims}}

Our descent procedure for the 2d and 4d non-Abelian conformal anomaly is easily generalized to $2n$ dimensions. The Euler invariant polynomial for $SO(2n+1,1)$ is given in \eqref{eq:gen Euler}. Its corresponding CS form is  
\begin{eqnarray}\label{eq:highercs}
Q^{\rm Euler}_{2n+1}=\frac{1}{(4\pi)^{n+1}n!}   \epsilon_{i_1\ldots i_{2n+2}}\,\int_0^1\,dt\,A^{i_1i_2}\wedge F_t^{i_3 i_4}\wedge\ldots\wedge F_t^{i_{2n+1}i_{2n+2}}\, ,
\end{eqnarray}
where $F_t=tdA+t^2A\wedge A$. Varying $Q^{\rm Euler}_{2n+1}$ with gauge parameter $\varepsilon=\varepsilon_{ij}T^{ij}$, we get $\delta_\varepsilon Q^{\rm Euler}_{2n+1}=dQ^{{\rm Euler}, 1}_{2n}$, where
 \begin{eqnarray}
Q^{{\rm Euler},1}_{2n}(\varepsilon,A)=\frac{1}{(4\pi)^{n+1}(n-1)!}\epsilon_{i_1\ldots i_{2n+2}} \left[\int_0^1 d t\,(1-t)\, \,\varepsilon^{i_1i_2} d\left(A^{i_3i_4}\wedge F^{i_5i_6}_t\wedge\ldots\wedge F_t^{i_{2n+1}i_{2n+2}}\right)\right]\,,\nonumber\\
\end{eqnarray}
which is the non-Abelian conformal anomaly in $2n$-dimensions. Note that the CS term \eqref{eq:highercs} allows to immediately write the \textit{anomaly theory} for the $2n$ dimensional conformal anomaly, via \textit{anomaly inflow} between the $2n+1$ dimensional bulk and its $2n$ dimensional boundary.

\section{Conclusions and future work\label{sec:conclusions}}

The main result of this paper is the derivation of the non-Abelian conformal anomaly in $2n$ dimensions via descent from the Euler invariant polynomial for $SO(2n+1,1)$ in $2n+2$ dimensions, and the resulting \textit{anomaly inflow} picture for the non-Abelian conformal anomaly. Our results indicate that the non-Abelian conformal anomaly is not exceptional, but admits the same cohomological descent and inflow interpretation as ordinary perturbative anomalies. We also compared and contrasted the $4d$ non-Abelian conformal anomaly with its type-A Weyl cousin, and showed how, while the latter solves a slightly different cohomology problem, can be obtained from a Weyl-cocycle part of the non-Abelian anomaly.

Furthermore, in 4d our descent procedure allowed us to write down a dilaton effective action realizing $SO(5,1)/ISO(4)$ and reproducing the full $SO(5,1)$ anomaly. The latter can be used for 't Hooft anomaly matching for the full $SO(5,1)$ anomaly, in the spirit of the Komargodski-Schwimmer dilaton effective action, which matches the anomaly under Weyl transformations. Note, however, that our dilaton effective action differs by construction from the KS action in that the former matches the non-Abelian anomaly while the latter reproduces its projection into a Weyl cocycle.

Our study leaves many desired directions to explore; first and foremost -- since our dilaton effective action differs from KS, we would like to study its implications for RGE flows. In other words, we wish to explore the constraints on RGE flows if one demands $SO(5,1)$ \textit{conformal} anomaly matching rather than \textit{Weyl anomaly matching}. This would involve studying positivity bounds in our WZW action in search of an $SO(5,1)$ version, rather than a Weyl version, of the a-theorem. In particular we will explore the interplay between $\Delta a$, the dilaton decay constant $f$, and the cutoff $\Lambda$ for the IR theory. We note that our conformal anomaly presents an obstruction to gauging translations even in a torsionless background, due to the background contribution of $F(A_f)\wedge F(A_\omega)$, the field strengths in the ``special conformal" and ``Lorentz" directions. Nevertheless we expect no obstruction to gauge \textit{diffeomorphisms}, namely local deformations of the vielbein supplemented by non-gauge transformations maintaining the Cartan constraints.

Needless to say, this work only outlined SZ descent for the non-Abelian anomaly, and its projection to a Weyl cocycle that recovers the type-A Weyl anomaly; we are currently seeking similar-in-spirit classifications for type-B Weyl anomalies, perhaps by seeking different solutions to the $SO(5,1)$ WZ consistency condition or alternatively, finding several different systematic ways to project the non-Abelian anomaly into Weyl cocycles. 

In this work, we argued that the non-Abelian conformal anomaly in Euclidean space is not quantized because the Euler invariant polynomial does not lift to an integer-valued characteristic class. Colloquially, we would say ``there are no 6d $SO(5,1)$ instantons". We would like to explore the analogous story in Lorentzian signature, where, naively, the Euler invariant polynomial for $SO(4,2)$ does define a bona-fide characteristic class.

Finally, our anomaly-matching approach involved treating the conformal currents as being manifestly independent, similar to the usual currents associated with any other non-Abelian symmetry. This is in contrast with the usual Riemannian-background construction of conformal currents from conformal Killing vectors and the energy momentum tensor. We suspect that this common-lore view of inter-dependent conformal currents may be an artifact of considering a Riemannian background, and that in fact under more general backgrounds, conformal currents should be truly independent. A future litmus test for that would be the explicit construction of a manifestly (classically) conformally invariant theory in a general, torsionful $SO(5,1)$ background -- for example by nonlinearly realizing $SO(5,1)$ using $U$ from Section~\ref{sec:dilaton}. We plan to study the conformal currents, and the general $SO(5,1)$ anomaly of such a model, in future work.

\section{Acknowledgements}
We thank Yael Shadmi for early collaboration on this project. We thank the JHEP referee for their insightful comments. We also thank Kai Bartnick, Shmuel Elitzur, Tom Hartman, Arthur Platschorre, Jake Solomon, Misha Smolkin, and Stefan Theisen for useful discussions. We thank Adam Schwimmer for comments on the manuscript and for probing us about the $SU^*(4)$ formulation of the anomaly. A special thanks to Zohar Komargodski for his invaluable comments on the draft and his questions regarding the dependence/independence of conformal currents. 
Part of the work was performed at the Aspen Center for Physics, which is supported by the NSF grant PHY-2210452.
CC is supported in part by the NSF grant PHY-2309456 and in part by the US-Israeli BSF grant 2024091. 
OT and GA are supported by the ISF grant
No. 3533/24, by the NSF-BSF grant No. 2022713, and by the BSF grant No. 2024169. 
SY was supported in part by a grant 01034816 titled ''String theory reloaded- from fundamental questions to applications" of the Planning and budgeting committee.

\appendix
\section{2d conformal anomaly from left and right movers}\label{app:SU2}
In Section~\ref{sec:desc} we showed that the 2d conformal anomaly descends from the 4d Euler invariant polynomial for $SO(4)$ (more accurately $SO(3,1)$), while the 2d gravitational anomaly descends from the 4d Pontryagin invariant polynomial. Here we touch base with a different partition of these invariant polynomial in terms of the Chern characters of $SU_L(2)\times SU_R(2)\cong SO(4)$. We use 't Hooft symbols \cite{tHooft1976} to linearly map the $SO(3,1)$ (anti-hermitian) generators $T^{ab},\,a,b\in\{1,\dots,4\}$ and the (anti-hermitian) basis $T^i_L,\,T^i_R,\,i\in\{1,\dots,3\}$ of $SU_L(2)\times SU_R(2)$:
\begin{eqnarray}
T^{i}_L=\frac{1}{2}\eta^i_{ab}T^{ab}~~,~~T^{i}_R=\frac{1}{2}\bar{\eta}^i_{ab}T^{ab}\,,
\end{eqnarray}
where
\begin{eqnarray}
&&\eta_{0 j}^i=i\delta_j^i~~,~~\eta_{j k}^i=\epsilon_{i j k} \nonumber\\[5pt]
&&\bar{\eta}_{0 j}^i=-i\delta_j^i~~,~~\bar{\eta}_{j k}^i=\epsilon_{i j k}\,,
\end{eqnarray}
and $\eta^i_{ba}=-\eta^i_{ab},\,\bar{\eta}^i_{ba}=-\bar{\eta}^i_{ab}$. We then have
\begin{eqnarray}
[T^i_L,T^j_L]=\epsilon_{ijk}T^k_L~~,~~[T^i_R,T^j_R]=-\epsilon_{ijk}T^k_R\,.
\end{eqnarray}
We can now rexpress the $SO(3,1)$ gauge field as
\begin{eqnarray}
&&A=\frac{1}{2}A^{ab}T^{ab}=A_L+A_R\nonumber\\[5pt]
&&A_{L,R}=A^i_{L,R}T^i_{L,R}\,.
\end{eqnarray}
Writing the Chern characters for $A_{L,R}$, we get
\begin{eqnarray}
I^{\rm Chern}_4(A_{L,R})=-\frac{1}{8\pi^2}{\rm tr}\left(F_{L,R}\wedge F_{L,R} \right)\,.
\end{eqnarray}
We explicitly checked that
\begin{eqnarray}
I^{\rm Euler}_4(A)&=&\frac{i}{4}\left[I^{\rm Chern}_4(A_{L})-I^{\rm Chern}_4(A_{R})\right]\nonumber\\[5pt]
I^{\rm Pontryagin}_4(A)&=&I^{\rm Chern}_4(A_{L})+I^{\rm Chern}_4(A_{R})\,.
\end{eqnarray}
Note the important factor of $i$ here in front the Euler invaiant polynomial. It is tied to the fact that the conformal anomaly is real, while chiral anomalies are a phase. Since a 2d right mover contributes to the $SU_R(2)$ anomaly with an opposite sign to the contribution of a left mover to the $SU_L(2)$ anomaly, we reproduce the well known fact \cite{AlvarezGaumeWitten1984,AlvarezGaumeGinsparg1985,Chamseddine:1992ry} that the gravitaional and Weyl anomalies correspond to the difference and sum of the left mover and right mover, respectively.

\section{4d non-Abelian conformal anomaly from the Chern character of $SU^*(4)$}\label{app:SU4}
The lie group $SO(6)$ is famously isomorphic to $SU(4)$. Analytically continuing to $SO(5,1)$ gives us the algebra $SU^*(4)\cong SO(5,1)$. Via this isomorphism, the Euler invariant polynomial for $SO(5,1)$ is proportional to the 3rd Chern character\footnote{More correctly, for $SU^*(4)$ it is an invariant polynomial that does not lift to an integral characteristic class, in contrast with the usual $SU(4)$ case.} of $SU^*(4)$, the noncompact real form of $SU(4)$. Hence we are free to consider the SZ descent from the 6d 3rd Chern character of $SU^*(4)$ to get the non-Abelian conformal anomaly. Specifically, 
\begin{eqnarray}\label{eq:chernEu}
I^{\rm Chern}_{6}=-\frac{i}{48 \pi^{3}}  {\rm tr}\left(F^3\right)\, ,
\end{eqnarray}
and the descent gives
\begin{eqnarray}
Q^{\rm Chern,1}_{4}(\varepsilon)=-\frac{i}{48 \pi^{3}}  {\rm tr}\,d\left[\varepsilon A\wedge dA+\frac{1}{2}A\wedge A^2\right]\, .
\end{eqnarray}
We checked explicitly that the identification of the conformal generators in $SU^*(4)$ gives $d^{DM_{ab}M_{cd}}=\epsilon_{abcd}$, and so we reproduce the non-Abelian conformal anomaly -- up to an overall factor. This factor decidedly involves an $i$, which is what we need to convert from a chiral anomaly to the (real) conformal anomaly.

Finally, one may worry that the descent from the Chern character \eqref{eq:chernEu} resembles the standard descent of a chiral anomaly, superficially at odds with the parity-even result $E_4$. This logic would have been correct had we taken $SU^*(4)$ to be an \textit{internal symmetry}. Here instead $SU^*(4)\cong SO(5,1)$ is a spacetime symmetry and so parity acts as a nontrivial automorphism on its generators, under which $P^a,\,K^a,\,M^{1a},\,a=2,3,4$ are odd. This guarantees that $d^{DM_{ab}M_{cd}}=\epsilon_{abcd}$ is parity odd, ensuring together with the implicit $\epsilon_{\mu\nu\rho\sigma}$ that the overall anomaly is parity-even. Lastly, the anomaly isn't quantized because there are no instantons for $SU^*(4)\cong SO(5,1)$.

\section{Gauged WZW Term}\label{app:WZW}
Gauged WZW have a long history in the literature, see for example \cite{Wess:1971yu,Witten:1983ar,Zumino:1983ew,Wu:1984pv,Hull:1990ms,DHoker:1984ph,DHoker:1994rdl}. Here we follow the notation of \cite{Brauner:2018zwr}. We take $G=SO(5,1)$ and $H=ISO(4)$. The coset $U\in SO(5,1)/ISO(4)$ is implicitly parametrized by \eqref{eq:coset} with the inverse Higgs relation \eqref{eq:IH}, but we are agnostic to this parametrization in this appendix. Following \cite{Brauner:2018zwr} we define:
\begin{eqnarray}
\theta\equiv U^{-1}dU~~,~~\overline{\theta}\equiv\theta+U^{-1}AU\equiv \theta+\overline{A}\,,
\end{eqnarray}
namely, $\theta$ is the Maurer-Cartan form, while $\overline{\theta}$ is $A$ gauge transformed by $U$. We also split $\theta,\,\overline{\theta}$ and $\bar{A}$ into projections along the broken generators $K^a,\,D$ and the unbroken generators $P^a,\,M^{ab}$ as follows:
\begin{eqnarray}\label{eq:defphi}
\phi=\theta^K_a K^a+\theta^D D~~&,&~~V=\theta^P_a P^a+\frac{1}{2}\theta^M_{ab} M^{ab}\nonumber\\[5pt]
\overline{\phi}=\overline{\theta}^K_a K^a+\overline{\theta}^D D~~&,&~~\overline{V}=\overline{\theta}^P_a P^a+\frac{1}{2}\overline{\theta}^M_{ab} M^{ab}\nonumber\\[5pt]
\overline{A}_\perp=\overline{A}^K_a K^a+\overline{A}^D D~~&,&~~\overline{A}_\parallel=\overline{A}^P_a P^a+\frac{1}{2}\overline{A}^M_{ab} M^{ab}\,.
\end{eqnarray}
Finally we define the field strengths
\begin{eqnarray}
F\equiv dA+A\wedge A~~&,&~~\overline{F}\equiv d\overline{\theta}+\overline{\theta}\wedge \overline{\theta}\nonumber\\[5pt]
W\equiv dV+V\wedge V~~&,&~~\overline{W}\equiv d\overline{V}+\overline{V}\wedge \overline{V}\,,
\end{eqnarray}
where the $SO(5,1)$ indices are implicitly contracted. The $ISO(4)$ invariant gauged WZW term for $SO(5,1)$ is then
\begin{eqnarray}\label{eq:WZWgen}
&&\Omega^{\rm Euler}_{5}\left(A,U\right)=Q^{\rm Euler}_5(A)-\nonumber\\[5pt]
&&\frac{1}{384  \pi^{3}}  \left\{\frac{1}{10} \etr{\overline{\phi}}{\overline{\phi}^2}{\overline{\phi}^2}-\frac{1}{2}\etr{\overline{W}+\overline{F}}{\overline{\phi}}{\overline{\phi}^2}+\right.\nonumber\\[5pt]
&&\left.\etr{\overline{W}}{\overline{W}}{\overline{\phi}}+\etr{\overline{F}}{\overline{F}}{\overline{\phi}}+\frac{1}{2}(\etr{\overline{W}}{\overline{F}}{\overline{\phi}}+\etr{\overline{F}}{\overline{W}}{\overline{\phi}}\right\} .
\end{eqnarray}
This construction guarantees that 
\begin{eqnarray}
d\Omega^{\rm Euler}_{5}\left(A,U\right)=0~~,~~\delta_\epsilon \Omega^{\rm Euler}_{5}\left(A,U\right)=dQ^{{\rm Euler},\,1}_{4}\left(\epsilon\right)\,.
\end{eqnarray}
The almost 4d form of this is \cite{Brauner:2018zwr}
\begin{eqnarray}\label{eq:WZW4d}
\Omega^{\rm Euler}_{5}\left(A,U\right)=\Omega^{\rm Euler}_{5}\left(0,U\right)+dB_4(A,U)\,,
\end{eqnarray}
with
\begin{eqnarray}\label{eq:B4}
&& B_4(A,U)=\left\langle\frac{1}{2} \phi^3\left(\bar{A}+\bar{A}_{\|}\right)+\frac{1}{4} \phi \bar{A}_{\perp} \phi\left(\bar{A}+\bar{A}_{\|}\right)+\frac{1}{2} \phi^2\left[\bar{A}_{\perp}, \bar{A}_{\|}\right]\right. \nonumber\\[5pt]
&& \quad+\phi\left(\frac{1}{2} \bar{A}_{\perp}^3+\frac{3}{4} \bar{A}_{\perp}^2 \bar{A}_{\|}+\frac{3}{4} \bar{A}_{\|} \bar{A}_{\perp}^2+\frac{1}{2} \bar{A}_{\|}^2 \bar{A}_{\perp}+\frac{1}{2} \bar{A}_{\|} \bar{A}_{\perp} \bar{A}_{\|}+\frac{1}{2} \bar{A}_{\perp} \bar{A}_{\|}^2+\bar{A}_{\|}^3\right) \nonumber\\[5pt]
&& \quad+\frac{1}{2} \bar{A}_{\perp} \bar{A}_{\|}^3-\frac{1}{2} \bar{A}_{\|} \bar{A}_{\perp}^3-\frac{1}{4} \bar{A}_{\|} \bar{A}_{\perp} \bar{A}_{\|} \bar{A}_{\perp} \nonumber\\[5pt]
&& \left.+\frac{1}{2} \bar{F}\left[\bar{A}+\frac{1}{2} \bar{A}_{\|}, \phi\right]+\frac{1}{2}(\bar{W}+W)\left[\frac{1}{2} \bar{A}+\bar{A}_{\|}, \phi\right]+\left(\frac{1}{2} \bar{F}+\frac{1}{2} \bar{W}+\frac{1}{4} W\right)\left[\bar{A}_{\|}, \bar{A}_{\perp}\right]\right\rangle\,.\nonumber\\
\end{eqnarray}
Note the slightly abuse of notation in this expression, as we didn't clearly parse it into the $\left\langle  X,Y,Z\right\rangle$ structure to make the gauge contractions manifest. As we do not make use of this explicit form in this paper, we omit this detail.

\providecommand{\href}[2]{#2}\begingroup\raggedright\endgroup

\end{document}